%% file: main.tex
\documentclass[%
 reprint,
 amsmath,amssymb,
 aps,
prb,
]{revtex4-2}

\usepackage{graphicx}
\usepackage{dcolumn}
\usepackage{bm}
\usepackage{braket}
\usepackage{amsmath}

\usepackage{xcolor}
\newcommand{\nv}{{N$V$}}
\newcommand{\nvm}{{N$V^-$}}
\newcommand{\nvz}{{N$V^0$}}

\newcommand{\nsp}{{N$_{\rm s}^+$}}

\newlength{\figwidth}
\begin{document}

\title{Spectroscopy of photoionization from the $^1E$ singlet state in nitrogen--vacancy centers in diamond}

\author{Sean M. Blakley}
\email{smb784@umd.edu}
\author{Thuc T. Mai}
\author{Stephen J. Moxim}
\author{Jason T. Ryan}
\author{Adam J. Biacchi}
\author{Angela R. Hight Walker}
\author{Robert D. McMichael}
\email{robert.mcmichael@nist.gov}
\affiliation{National Institute of Standards and Technology, Gaithersburg, MD 20899 USA}

\date{\today}

\begin{abstract}
    The $^1E$---$^1A_1$ singlet manifold of the negatively charged nitrogen vacancy (\nvm) center in diamond plays a central role in its quantum information and quantum sensing applications. However, the energy gap between the $^1E$ singlet state and diamond band edges or the $^3A_2$---$^3E$ triplet manifold has not been measured directly. Using field-quenching effects on photoluminescence (PL) spectra, we measure these energy gaps as a function of temperature, applied magnetic field, excitation wavelength, and excitation power in a heavily nitrogen-doped sample. Increased PL and decreased zero-phonon line width from the \nvz\ were observed in the presence of an applied magnetic field, indicating ionization from the long-lived $^1E$ singlet state. A temperature-dependent ionization threshold between 532~nm and~550 nm was found, locating the singlet states within the diamond band gap.
\end{abstract}

\maketitle

\section{Introduction\label{sec:introduction}}
\input{introduction.tex}

\section{Methods\label{sec:methods}}

\input{methods.tex}

\section{Results \& Discussion\label{sec:results}}
\input{results.tex}

\section{Conclusion\label{sec:conclusion}}
\input{conclusion.tex}

\begin{acknowledgements} 
We wish to acknowledge Ilya Fedotov and Masfer Alkahtani for many helpful discussions and invaluable advice, Veronika Szalai for initial EPR measurements, and Adam Gali for insightful comments on the manuscript.
\end{acknowledgements}

\bibliography{main}

\end{document}


\title{Supplementary Material: Effect of singlet state ionization on photoluminescence quenching from nitrogen vacancy centers in diamond}

\author{Sean M. Blakley}
\email{sean.blakley@nist.gov}
\author{Thuc Mai}
\author{Stephen J. Moxim}
\author{Jason T. Ryan}
\author{Adam J. Biacchi}
\author{Angela Hight-Walker}
\author{R. D. McMichael}
\email{robert.mcmichael@nist.gov}
\affiliation{National Institute of Standards and Technology, Gaithersburg, MD 20899 USA}

\date{\today}

\begin{abstract}
The nitrogen vacancy color center in diamond has proven to be a convenient platform for exploration of quantum effects and quantum measurements.
\end{abstract}

\maketitle

\section*{Bayesian Inference\label{sec:Bayesian Inference}}
\input{analysis.tex}

\section*{Photodynamics\label{sec:Photodynamics}}
\input{photodynamics.tex}

\section*{Surface Plots for \nvz\ Contrast Parameters\label{sec:SurfacePlots}}

Fig. \ref{fig:alphadelta} contains surface plot reproductions of Fig. 5 in the manuscript.

\begin{figure}[hbt!]
    \centering
    \includegraphics[width=\figwidth]{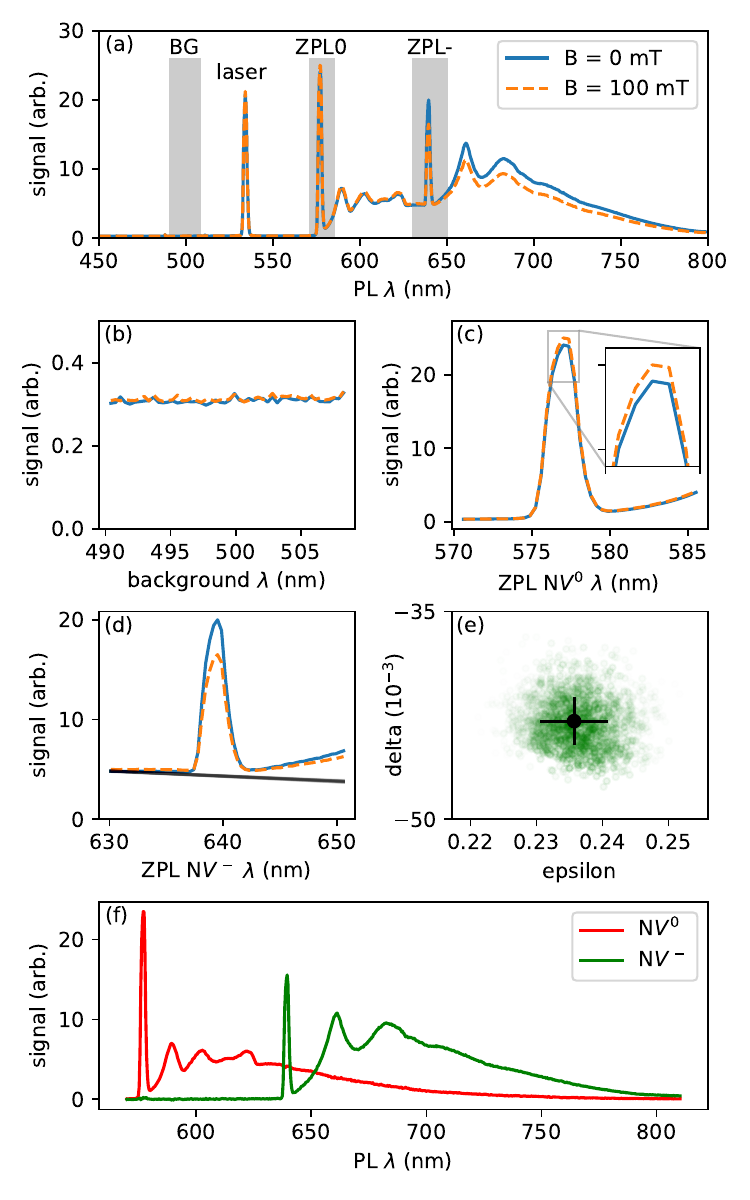}
    \caption{Example spectra illustrating Bayesian inference methods. (a) Raw spectra collected with 40 $\mu$W, 532 nm excitation at 1.6 K. The labeled grey rectangles indicate zones where data is used in Bayesian inference. The legend applies to panels (a-e). (b) Constant background in zone BG. (c) Spectra in zone ZPL0 encompassing the zero-phonon line of \nvz. The inset highlights an increase in photoluminescence with applied field. (d) Spectra in zone ZPL- encompassing the zero-phonon line of \nvm. The black line is the inferred $I^0$ spectrum assuming linear wavelength dependence.  The signal decreases with applied field. (e) Particle swarm representation of $P(\epsilon, \delta | {\bm I}_{0}, {\bm I}_{B}, {\bm \lambda})$ (f) Separated contributions, $I^-$ and $I^0$ calculated using the mean values of $\epsilon$ and $\delta$. The absence of a \nvm peak in the $I_0$ spectrum and the absence of a \nvz peak in the $I^-$ spectrum indicate good quality of the inference results.}
    \label{fig:fitexample2}
\end{figure}

\begin{figure*}[hbt!]
    \centering
    \includegraphics[width=\textwidth]{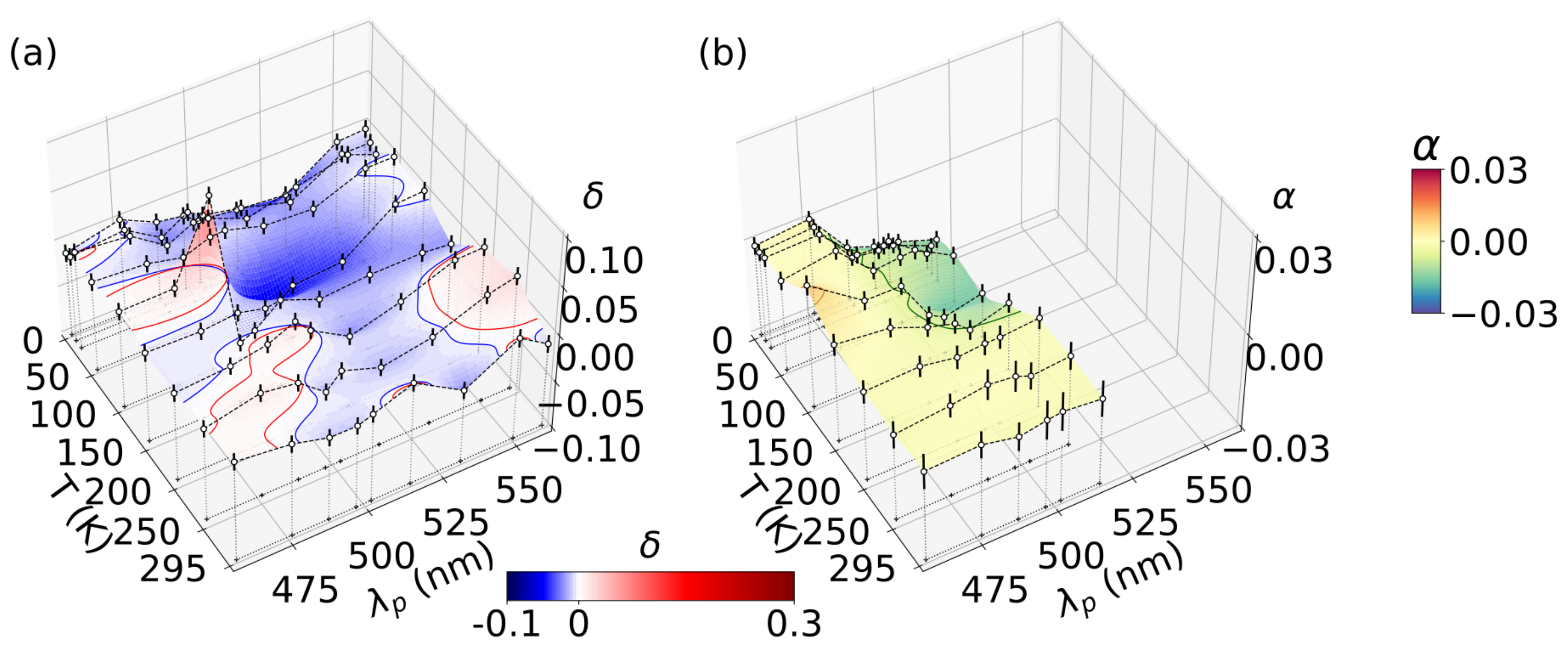}
    \caption{Field-induced contrast parameters for \nvz\ PL. (a) The error-weighted power average of \nvz\ PL contrast $\delta$; negative values in blue indicate increasing PL contrast with field. (b) The \nvz\ ZPL line broadening parameter, $\alpha$, for 40 $\mathrm{\mu} W$ laser power.  Negative values in green indicating narrowing under applied field.  Solid red, blue contours in (a) and orange, green contours in (b) mark two standard deviations above and below zero contrast respectively.  Color surfaces represent bivariate spline interpolation between measured data points (white circles with error bars). Black dashed lines connecting data points serve to aid the eye.}
    \label{fig:alphadelta}
\end{figure*}

\bibliography{supp}

%% file: introduction.tex
The nitrogen vacancy (\nv) center in diamond has emerged as an important platform for quantum technology.\cite{aharonovich_diamond_2014, awschalom_quantum_2018, barry_sensitivity_2020, bernardi_nanoscale_2017, bourgeois_photoelectric_2020, doherty_nitrogen-vacancy_2013, gali_recent_2023, jelezko_single_2006, rondin_magnetometry_2014, schirhagl_nitrogen-vacancy_2014, schroder_quantum_2016, wrachtrup_single_2016} The negative charged state (\nvm) garners the most interest due to its $S=1$ triplet ground state. These ground state sublevels have long spin coherence times and can be polarized and read out optically at room temperature.\cite{balasubramanian_ultralong_2009, yamamoto_extending_2013}. 

\begin{figure}
    \centering
    \includegraphics[width=\figwidth]{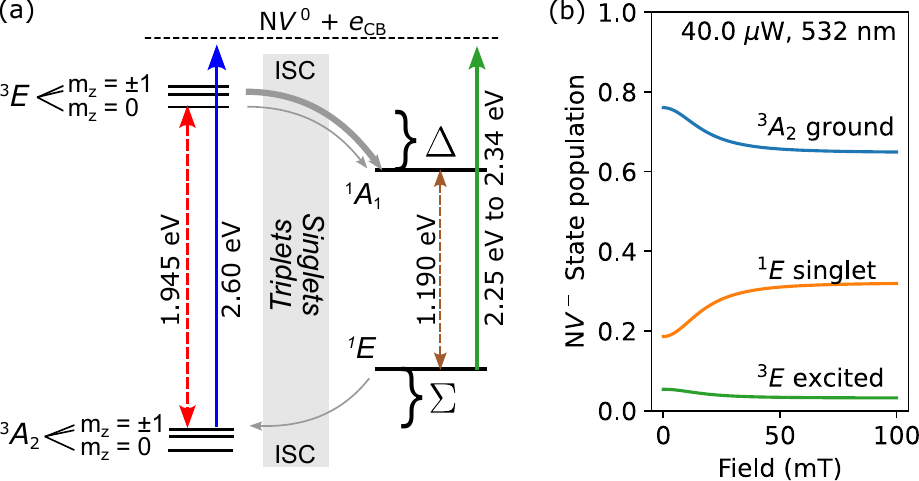}
    \caption{(a) Diagram of energy levels and transitions in the \nvm\ center. Energy differences corresponding to zero phonon lines are indicated by dashed arrows, and ionization transitions are shown with solid arrows. The ISC across the $^3E$---$^1A_1$ energy gap $\Delta$ is spin dependent and is primarily responsible for the optically detected magnetic resonance (ODMR) properties of the \nvm\ centers. The bounds placed on the ionization threshold of the singlet state are a main result of this work. (b) Modeled state populations under constant illumination as a function of field applied along the diamond [001] axis. Increasing population of the $^1E$ singlet state with applied field is used to identify $^1E$ ionization.}
    \label{fig:levels_and_fieldplot}
\end{figure}

Initialization and readout of the \nvm\ spin state involves transitions between four electronic states within the diamond band gap: the S = 1 spin triplet ground ($^3A_2$) and excited ($^3E$) electronic states, and the S = 0 spin singlet metastable ($^1E$) and excited ($^1A_1$) electronic states (Fig. \ref{fig:levels_and_fieldplot}(a)), which we refer to as triplet and singlet states respectively. For illumination wavelengths between 470 nm and 637 nm, \nvm\ centers will absorb photons and transition from ground to the excited $^3E$ states while preserving $m_z$. Relaxation back to the ground state can occur directly by emitting a photon, or indirectly via the inter-system crossings (ISCs) and a transition between singlet states, typically without emitting a photon (Fig. \ref{fig:levels_and_fieldplot}(a)). Importantly, the ISC path is more probable for $m_z = \pm 1$ than for $m_z=0$.\cite{doherty_nitrogen-vacancy_2013} The spin-dependence of the upper ISC makes it possible to prepare the $m_z = 0$ spin state with good fidelity and to read it out via photoluminescence (PL). The lower ISC is a slow process that gives the metastable $^1E$ state a lifetime that is an order of magnitude longer than any other excited state.\cite{doherty_nitrogen-vacancy_2013}

The $^3A_2$---$^3E$ and $^1E$---$^1A_1$ transition energies in the \nvm\ are determined by the photon energies of the zero phonon lines (ZPL) present in the absorption and PL spectra near 1.95~eV and 1.190~eV respectively.\cite{beha_optimum_2012,doherty_nitrogen-vacancy_2013,thiering_theory_2018,rogers_infrared_2008,acosta_optical_2010} The single-photon ionization and recombination thresholds place the \nvm\ ground state 2.60~eV below the conduction band and 2.94 eV above the valence band.\cite{aslam_photo-induced_2013}

Using ionization thresholds to place the $^1A_1$ and $^1E$ singlet states within the bandgap has proven more difficult. Previous studies have focused on measuring the $^1A_1$---$^3E$ and $^3A_2$---$^1E$ energy gaps, labeled $\Delta$ and $\Sigma$ respectively in Fig. \ref{fig:levels_and_fieldplot} (a). Estimates of $\Delta$ at cryogenic temperatures and temperatures above 300~K have been calculated by relating ISC rates to these energy gaps, and values of $\Sigma$ have been inferred using these calculations for $\Delta$ and the ZPL energy of the $^1E$---$^1A_1$ transition without investigating $\Sigma$ directly. At cryogenic temperatures, $\Delta$ is estimated to lie between 0.321~eV and 0.414~eV,\cite{goldman_phonon-induced_2015, goldman_state-selective_2015, goldman_erratum_2017} and earlier work done at high temperature is at variance with these results, yielding an estimate of $\Delta \approx 0.8$~eV.\cite{toyli_measurement_2012} These differing results suggest a temperature dependent phenomena inherent to the $^1A_1$---$^1E$ singlet manifold. This is the subject of investigation in this work.

Often these values of $\Delta$ are used in conjunction with the ZPL energy of the $^1E$---$^1A_1$ transition and the ionization energy of the $^3A_2$ to locate the $^1E$ with respect to the diamond conduction band edge, however a recent publication indicates that this common technique is inappropriate for accurately determining the $^1E$ ionization energy as it erroneously treats the inherently multi-particle energy levels of the \nvm as single-particle energy levels.\cite{gali_recent_2023}  In light of this new information, the $^1E$ ionization energy must be directly measured, not inferred from existing data. 
To accomplish this, the $^1E$ state can be selectively populated using a magnetic field to mix the $m_z = 0$ and $m_z = -1$ triplet sublevels while under laser excitation, and then photoionized with wavelength tuned laser excitation, which would allow the ionization energy of the $^1E$ to be measured.  This technique can be applied in a temperature controlled cryostat in order to determine the $^1E$ ionization energy as a function of temperature.

Ionization of \nvm\ results in an increase of \nvz\ concentration and a corresponding increase in \nvz\ PL.  The \nvm\ system can be polarized into the $^1E$ by exploiting a magnetic field\cite{aude_craik_microwave-assisted_2020} or microwave resonance to increase the population of the $^3E$ $m_z = \pm 1$ states, which can then relax to  of $^1E$. The $^1E$ can then be selectively ionized by applying laser excitation to the system and reading out the change in \nvz\ PL.\cite{wolf_nitrogen-vacancy_2022,aude_craik_microwave-assisted_2020,hopper_near-infrared-assisted_2016,hacquebard_charge-state_2018}

In the present work, we use field modulation to investigate ionization dynamics as a function of temperature, wavelength, and optical power. The diamond has a high concentration of \nv\ centers suitable for ensemble measurements of interest to this work. As shown in Fig. \ref{fig:levels_and_fieldplot}(b), population transfers from the triplet ground to the $^1E$ singlet state when laser excitation is applied in the presence of a magnetic field. We exploit the fact that the PL spectra of \nvm\ and \nvz\ behave differently under the influence of a magnetic field to track their relative defect populations and identify ionization thresholds. We find that in the absence of a magnetic field, the fraction of \nv\ centers in the \nvm\ charge state is approximately constant for photon energies lower than a threshold between 2.54~eV and 2.60~eV, which we identify as the ionization energy of the $^3A_2$ ground state.\cite{aslam_photo-induced_2013} At energies below that threshold, the concentration of \nvz\ increases by a few percent when a spin-mixing magnetic field is applied,  due to increased ionization from the newly populated $^1E$ state. A threshold for the \nvz\ contrast between 532~nm and 550~nm brackets the singlet state ionization energy between 2.25~eV and 2.33~eV for temperatures between 1.6~K and 300~K.


%% file: methods.tex
\paragraph*{Materials and Methods: Experimental\label{sec:experimental}}


The sample is a commercial CVD-grown diamond (3 mm $\times$ 3 mm $\times$ 0.5 mm) with a typical substitutional nitrogen (N$_s$) fraction $\approx 13 \times 10^{-6}$ (13~ppm) and \nv\ fraction $\approx 4.5 \times 10^{-6}$ (4.5~ppm). Our quantitative electron spin resonance (ESR) measurements\cite{eaton_quantitative_2010} determined the ratio of ESR---active N$_s^0$:\nvm\ centers was approximately 3.1:1. ESR results also revealed the presence of N$V$H$^-$ centers in concentrations similar to \nv\ (\nvm:N$V$H$^-$ = 1.1:1), in agreement with extensive characterization of a similar sample.\cite{edmonds_characterisation_2021} Relatively small responses from additional defect centers were too heavily obscured by the central N$_s^0$ and N$V$H$^-$ resonances for a complete analysis. Absence of \nvz\ responses agrees with previous measurements\cite{felton_electron_2008} and may be explained by strain broadening.\cite{barson_fine_2019} Fitting to ESR simulations of the defects, when necessary, was done with EasySpin software\cite{stoll_easyspin_2006} using spin-system parameters taken from several sources.\cite{loubser_electron_1978,smith_electron-spin_1959,glover_hydrogen_2003} The quantitative ESR measurements contain at most a 15 \% error.

\begin{figure}
    \centering
    \includegraphics[width=\figwidth]{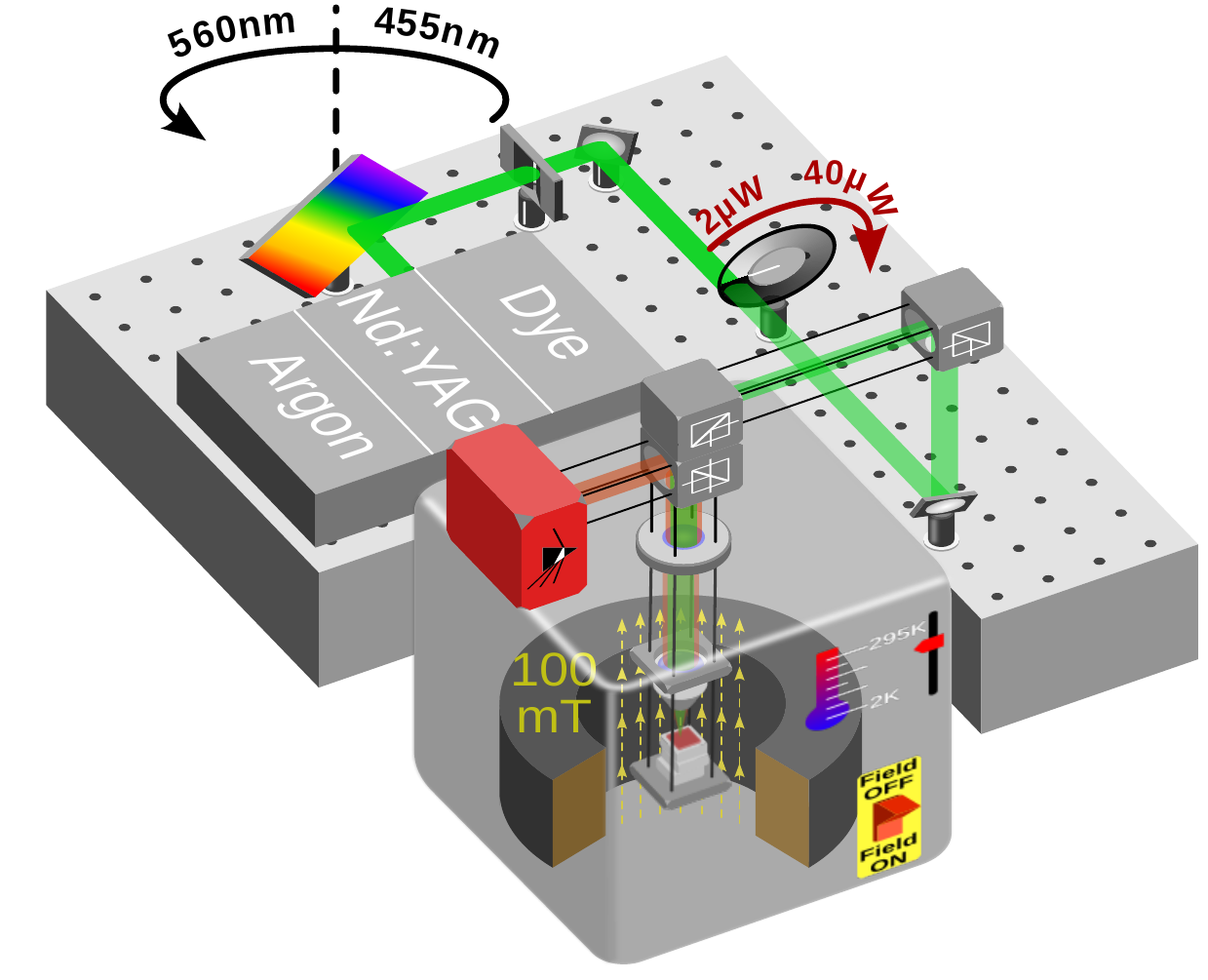}
    \caption{Diagram of the experimental setup.}
    \label{fig:exp_apparatus}
\end{figure}

The apparatus for photoluminescence measurement is illustrated in Fig. \ref{fig:exp_apparatus}. The sample is mounted inside a closed-loop, cryogen-free cryostat with an optical window and superconducting magnet. The cryostat is capable of set point temperatures between 1.6~K and 295~K, and perpendicularly applied magnetic field magnitudes from 0~T to 9~T. The sample is oriented with the applied field parallel to the [100] crystal axis of the diamond, $\approx 55^\circ$ away from the N-$V$ axes along the [111] direction.  Three different continuous wave (CW) lasers are used as a source of excitation for the PL: an argon ion laser which generates wavelengths at 458~nm, 476~nm, 488~nm, 497~nm, 502~nm, and 515~nm; a solid state Nd:YAG laser at 532~nm; and a dye laser at 550~nm and 560~nm. Each laser is directed into the cryostat and onto the sample using free space optics. The beam is at normal incidence to the [100] sample surface and focused through a 50$\times$ cryogenic objective. The emitted light is collected and sent to two single grating spectrometers, one with a large bandwidth and one with a higher spectral resolution. The laser spot size on the sample is approximately 1~$\mu$m. A continuous variable neutral density filter was used to adjust the power entering the optical window of the cryostat.  The optical power was measured by a power meter after the neutral density filter but prior to insertion into the cryostat chamber. 

\begin{figure*}
    \centering
    \includegraphics{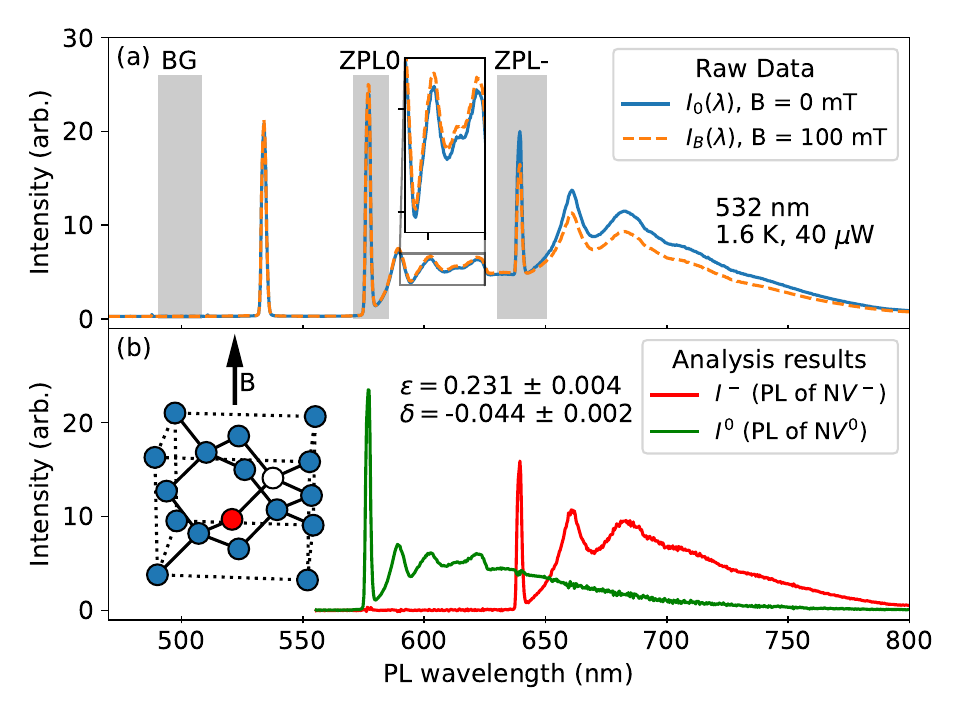}
    \caption{(a) Example raw spectra showing the combined photoluminescence of \nvz and \nvm at applied fields of 0~mT and 100~mT. The peak at $\approx$ 532~nm is the excitation laser. The inset magnifies a segment of the spectrum dominated by PL from \nvz, which exhibits contrast $\delta$ with applied field. The tail above 650~nm is dominated by PL from \nvm, which exhibits contrast $\epsilon$ with field. (b) Analysis results of the data in (a) showing the separated spectra $I^0$ and $I^-$ of \nvz\ and \nvm\ respectively, and also the inferred contrast values $\epsilon$ and $\delta$. Inset, the diamond crystal structure with an NV center: C (blue circles), N (red circle), vacancy (white circle). The magnetic field is along the [001] direction, at an angle of $\approx$ 55$^\circ$ to the N--$V$ axis, which is along the [111] direction.}
    \label{fig:fitexample}
\end{figure*}

Photoluminescence spectra were collected under CW illumination with ($I_B(\lambda)$) and without ($I_0(\lambda)$) a 100~mT magnetic field. These spectrum pairs were collected for all combinations of nine temperatures from 1.6~K to 295~K, nine laser wavelengths from 458~nm to 560~nm, and five optical powers at 2.5 $\mu$W, 5 $\mu$W, 10 $\mu$W, 20 $\mu$W, and 40~$\mu$W for a total of 405 spectrum pairs. An example spectrum pair is plotted in Fig. \ref{fig:fitexample}(a). Fig. \ref{fig:fitexample}(a) shows PL spectra from an ensemble of \nv\ centers with the separated contributions of \nvz\ and \nvm\ in Fig. \ref{fig:fitexample}(b). Each spectrum consists of a narrow, zero-phonon line (ZPL) and several broad features in the phonon sideband (Fig. \ref{fig:fitexample}(b)).

\paragraph*{Materials and Methods: Data Analysis\label{sec:data_analysis}}

Spectra collected from samples with large numbers of \nv\ centers have high enough signal-to-noise (SNR) to manipulate $I_0(\lambda)$ and $I_B(\lambda)$ and extract the \nvm\ and \nvz\ contributions with least-squares fitting\cite{aude_craik_microwave-assisted_2020, chakraborty_magnetic-field-assisted_2022}. Our implementation of these methods was not reliable enough for automated analysis.

To tackle the analysis of these spectrum pairs efficiently, we required a reliable technique that can be automated with a minimum of human intervention. We addressed this requirement by developing a Bayesian analysis that separates the contributions of the \nvz\ and \nvm\ through their differing responses to changes field.\cite{aude_craik_microwave-assisted_2020, chakraborty_magnetic-field-assisted_2022} The field-off spectra are modeled with zero-field photoluminescence amplitude contributions ${I}^{-}(\lambda)$ and ${I}^{0}(\lambda)$ from \nvm\ and \nvz\ respectively.
\begin{equation}
    I_{0}(\lambda) = I^-(\lambda) + I^0(\lambda) + C + \eta_0,
    \label{eq:Imodela}
\end{equation}
where we have added a constant background, $C$ and a measurement noise, $\eta_0.$ For the field-on spectra, we introduce quenching parameters $\epsilon$ and $\delta$ to describe the fractional decrease in PL amplitude from the \nvm\ and \nvz\ respectively due to the spin-mixing effects of the applied field.
\begin{equation}
  I_{B}(\lambda) = (1-\epsilon)I^-(\lambda) + (1-\delta)I^0(\lambda) + C + \eta_B,
\label{eq:Imodelb}  
\end{equation}
including background $C$ and an uncorrelated noise $\eta_B$.
Negative values of $\epsilon$ or $\delta$ indicate PL amplitude that increases with applied field.

Briefly, the Bayesian analysis only uses data from parts of the spectra where additional assumptions can be incorporated, indicated by shaded regions in Fig. \ref{fig:fitexample}(a). We assume that region BG contains background $C$ only, that region ZPL0 contains only $I^0(\lambda)$, with $I^-(\lambda) = 0$.  In region ZPL-, we assume that $I^0(\lambda)$ is a linear function of $\lambda$ to distinguish it from the sharply peaked structure of $I^-(\lambda)$. The analysis also incorporates measurement noise in the likelihood calculations.  With these assumptions and inputs, the analysis yields probability distributions for $\epsilon$ and $\delta$, allowing Eqns. (1-2) to be inverted, yielding separate contributions $I^-(\lambda)$ and $I^0(\lambda)$. Fig. \ref{fig:fitexample}(b) shows the separated example spectra with the inferred values of $\epsilon$ and $\delta$. The procedure was repeated for 405 field on/off pairs of PL spectra, using plots of the separate contributions as visual checks of the analysis quality. Details of the procedure are provided in the supplemental material.

We also looked for changes in the full-width half-maximum (FWHM) of the \nvz\ ZPL using our high-resolution spectrometer, but we were limited to analyzing data for excitation wavelengths less than 515~nm due to scattered laser light breaking through the high-resolution spectrometer's edge-pass filter. In a first attempt, least-square fits of the \nvz\ ZPL to Lorentzian line shapes yielded line widths of approximately 0.25~nm up to 100~K, then increasing to 1.5~nm at 300~K. Systematic misfit between the Lorentzian model and the peak data led to inflated uncertainty estimates.  To avoid this problem, the field-induced line width contrast was estimated using a model-free approach that amounts to finding the transformation that best mapped the $B = 0$~mT data onto a cubic spline interpolation of the $B = 100$~mT data. The transformation included "horizontal" expansion of the $\lambda$ axis around the approximate peak center by a factor of $(1 + \alpha)$, translation of the $\lambda$ axis to correct the center value, and ``vertical'' scaling of the $I_0(\lambda)$ axis, where $\alpha$ is the fraction of measured broadening in the \nvz\ ZPL FWHM. For each data point, the difference between the scaled $B = 0$~mT data and the interpolation of the $B = 100$~mT data is essentially an error that determines the likelihood of the scaling parameters given the data values.

\paragraph*{Materials and Methods: Modeling}

The populations of the \nv\ states are calculated using a five-state model for \nvm\ and a two-state model for \nvz.\cite{robledo_spin_2011, tetienne_magnetic-field-dependent_2012, liaugaudas_luminescence_2012, beha_optimum_2012, hacquebard_charge-state_2018} The five-state model of \nvm\ includes transitions between the $m_z = 0$ and combined $m_z = \pm 1$ spin states of $^3A_2$ electronic state, the the $m_z = 0$ and combined $m_z = \pm 1$ spin states of the $^3E$ level and the long lived $^1E$ singlet state. The $^1A_1$ state is ignored because of its very short lifetime. The \nvz\ is modeled using its ground and excited states. We assume that ionization and recombination rates are slow compared to the internal dynamics of each center, so that the steady state populations of the centers' states are independent of the centers' concentrations. See supplemental material for details.

%% file: results.tex

%
%
\begin{figure}
    \centering
    \includegraphics[width=\figwidth]{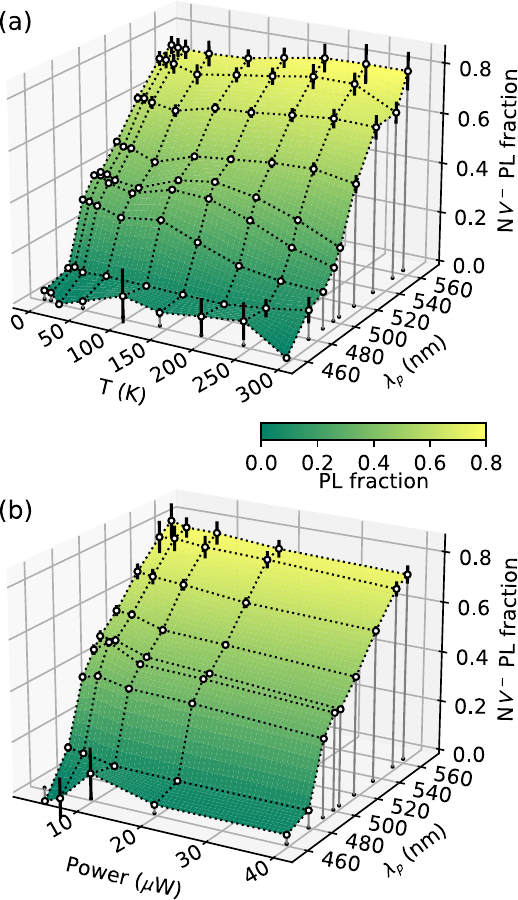}
    \caption{Fraction of photoluminescence due to \nvm. (a) PL fraction vs. temperature and wavelength at 40 $\mu$W laser power and (b) PL fraction as a function of power and wavelength at 1.6~K. The threshold between 476~nm and 488~nm excitation wavelengths is attributed to single-photon ionization of the $^3A_2$ ground state.}
    \label{fig:minusfrac}
\end{figure}

The relative PL amplitude of the \nvm\ and \nvz\ yields information about the charge state of \nv\ centers. The fraction of the integrated PL amplitude contributed by \nvm\ in zero applied field is plotted in Fig. \ref{fig:minusfrac}. Figs. \ref{fig:minusfrac}(a) and (b) both show a dropoff in \nvm PL amplitude between 476~nm and 488~nm (2.605~eV and 2.55~eV respectively). This threshold is consistent with ionization of \nvm\ from its ground state to \nvz\ with an optically  measured threshold near 2.6~eV (477 nm)\cite{aslam_photo-induced_2013}, and a photoconductivity threshold between 2.6~eV and 2.7~eV\cite{bourgeois_enhanced_2017}.  Calculated thresholds include 2.67~eV,\cite{razinkovas_photoionization_2021},  2.7~eV.\cite{bourgeois_enhanced_2017}, and 
2.76~eV\cite{bockstedte_supplementary_2018}. The threshold for ionization of the $^3A_2$ appears to shift to lower photon energies and/or broaden at the higher temperatures. A slight decreasing trend with power at all excitation wavelengths in Fig.\ 4\ref{fig:minusfrac}(b) shows that the \nvm\ PL fraction is only weakly power dependent.  We therefore consider only the error-weighted average over all powers when determining ionization thresholds from PL amplitudes as a function of temperature and wavelength.

%
%
We next compare the measured PL fraction with several models of charge dynamics, all of which assume that the integrated PL amplitude for \nvm\ and \nvz\ are proportional to the species concentrations $c_{NV^-}$, $c_{NV^0}$ and the excited state populations $n_{ex}^-(\lambda_p)$, and $n_{ex}^0(\lambda_p)$ respectively.
\begin{eqnarray}
\int I^-(\lambda)\, d\lambda & \propto & c_{NV^-} n_{ex}^-(\lambda_p) \\
\int I^0(\lambda)\, d\lambda & \propto & c_{NV^0} n_{ex}^0(\lambda_p).
\end{eqnarray}

A simple rate-balance photodynamic model of the steady state assumes that the rates of ionization and recombination must be equal,
\begin{equation}
    c_{NV^-}\Gamma_{NV^-} = c_{NV^0}\Gamma_{NV^0},
    \label{eq:steady_state}
\end{equation}
where $\Gamma_{NV^-}$ and $\Gamma_{NV^0}$ are the ionization rate of \nvm\ and  recombination rate of \nvz, respectively. For excitation wavelengths greater than 490~nm and powers on the order of 10~$\mu$W, the transition rates are quadratic in laser power and on the order of 10$^3$~s$^{-1}$.\cite{aslam_photo-induced_2013}. The transition rates are modeled as
\begin{eqnarray}
\Gamma_{NV^-} & = & J(n_{ex}^-\sigma_{ex}^- +n_s^-\sigma_s^- + n_g^-\sigma_g^-) \\
\Gamma_{NV^0} & = & J(n_{ex}^0\sigma_{ex}^0).
\end{eqnarray}
for photon flux $J$ and cross sections, $\sigma$. Superscripts indicate the charge state of the N$V$ center and subscripts $ex$, $s$ and $g$ refer to excited, singlet and ground states.  Because the populations of states other than the ground state are linear in power, the net ionization and recombination rates are quadratic. Ionization from the ground state is linear in power for excitation wavelengths shorter than 485~nm.

We also consider a fixed-concentration model, where the concentrations are only slightly perturbed by quadratic ionization and recombination processes, which agrees empirically with our observations.

\begin{figure}
    \centering
    \includegraphics[width=\figwidth]{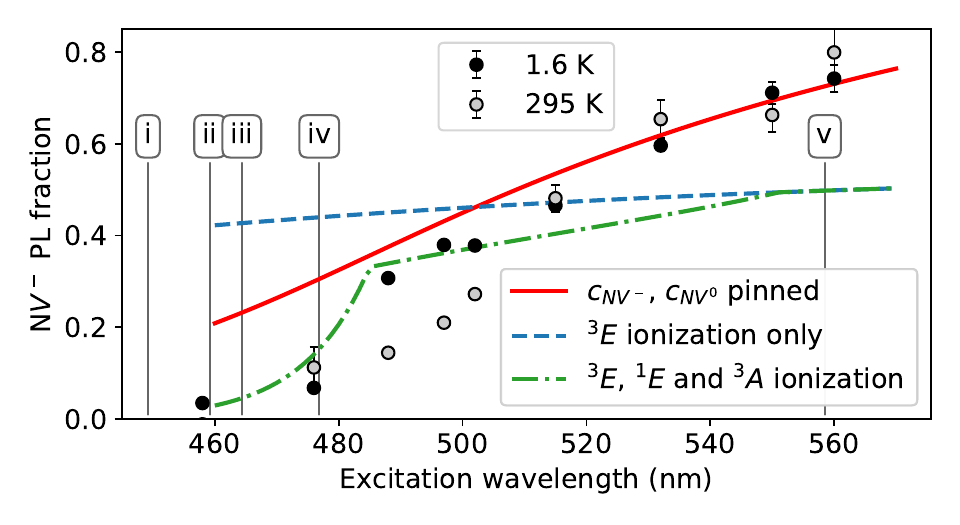}
    \caption{Fraction of integrated PL amplitude that is due to \nvm\ as a function of excitation wavelength. Measured data is from 1.6~K (black dots), 295~K (grey dots) with 40~$\mu$W excitation power. Error bars indicate one standard deviation. The curves are model results calculated for fixed charge concentrations (solid red line) and for concentrations inversely proportional to ionization rate: using only ionization from the $^3E$ excited state (dashed blue line) and using additional ionization from the $^1E$ singlet and the $^3A$ ground states (dot-dash green line). Modeled ionization thresholds are 551~nm (2.25~eV) from the singlet state and 485~nm for ionization from the ground state. See (Table 1) for description of the vertically marked wavelengths (i, ii, iii, iv, and v).
    }
    \label{fig:pl_models}
\end{figure}

\begin{table}
    \centering
    \begin{tabular}{c|c|c|c}
    Marker & Ion.\ energy & Init. state & Source \\
    \hline
         i & 2.76~eV & $^3A_2$
             & calculation\cite{bockstedte_supplementary_2018}\\
         ii & 2.7~eV &  $^3A_2$
             & calculation\cite{bourgeois_enhanced_2017}\\
         iii & 2.67~eV & $^3A_2$
             & calculation\cite{razinkovas_photoionization_2021}\\
         iv & 2.6~eV & $^3A_2$
             & measurement\cite{aslam_photo-induced_2013}\\
         v &  2.22~eV & $^1E$ 
             & calculation\cite{bockstedte_supplementary_2018}\\
    \end{tabular}
    \caption{Prior ionization threshold results at wavelengths marked in Fig.\ \ref{fig:pl_models}}.
    \label{tab:thresholds}
\end{table}
Fig.\ \ref{fig:pl_models} shows the behavior of the \nvm\ PL fraction using rate-balance and fixed-concentration models as a function of excitation wavelength. Measured results are provided for 1.6~K (black) and 295~K (grey fill) data. 

The blue dashed line in Fig.\ \ref{fig:pl_models} is the balanced rate model with ionization allowed only from the excited state. Under this assumption the \nvm\ PL fraction is nearly constant. The green dot-dash curve also includes ionization from the singlet state for wavelengths below 550~nm, and from the ground state below 485~nm. The solid red curve corresponds to the fixed-concentration model with $c_{NV^-} = 0.66$ and $c_{NV^0} = 0.34$, and the resulting \nvm\ PL fraction increases smoothly across the pump wavelength range. In view of the poorly determined parameter values, firm conclusions should not be drawn from these models. However, the results are in good agreement with previous measurements of two-step ionization/recombination rates on individual centers.\cite{aslam_photo-induced_2013}

Qualitative agreement between the fixed-concentration model results and measurement data suggests the \nvm\ and \nvz\ concentrations are approximately unchanged for excitation wavelengths longer than 490~nm in the absence of an external magnetic field. Figs.\ \ref{fig:minusfrac}, \ref{fig:pl_models} both suggest an ionization threshold from the $^3A_2$ ground state between 2.54~eV and 2.60~eV (488~nm and 476~nm resp.).

%
The effects of the applied magnetic field provide additional information about changes in charge state. The off axis field is known to decrease the population of the $^3E$ excited states, quenching the PL from \nvm. The field also increases the population in the $^1E$ singlet state.\cite{doherty_nitrogen-vacancy_2013,tetienne_magnetic-field-dependent_2012,chakraborty_magnetic-field-assisted_2022} By changing the populations, the applied field also affects net ionization rates of these states. The decrease in $^3E$ population and increase in $^1E$ population induced by the field could increase or decrease the total ionization rate, depending on their respective cross sections. 

%
%

\begin{figure}
    \centering
    \includegraphics[width=\figwidth]{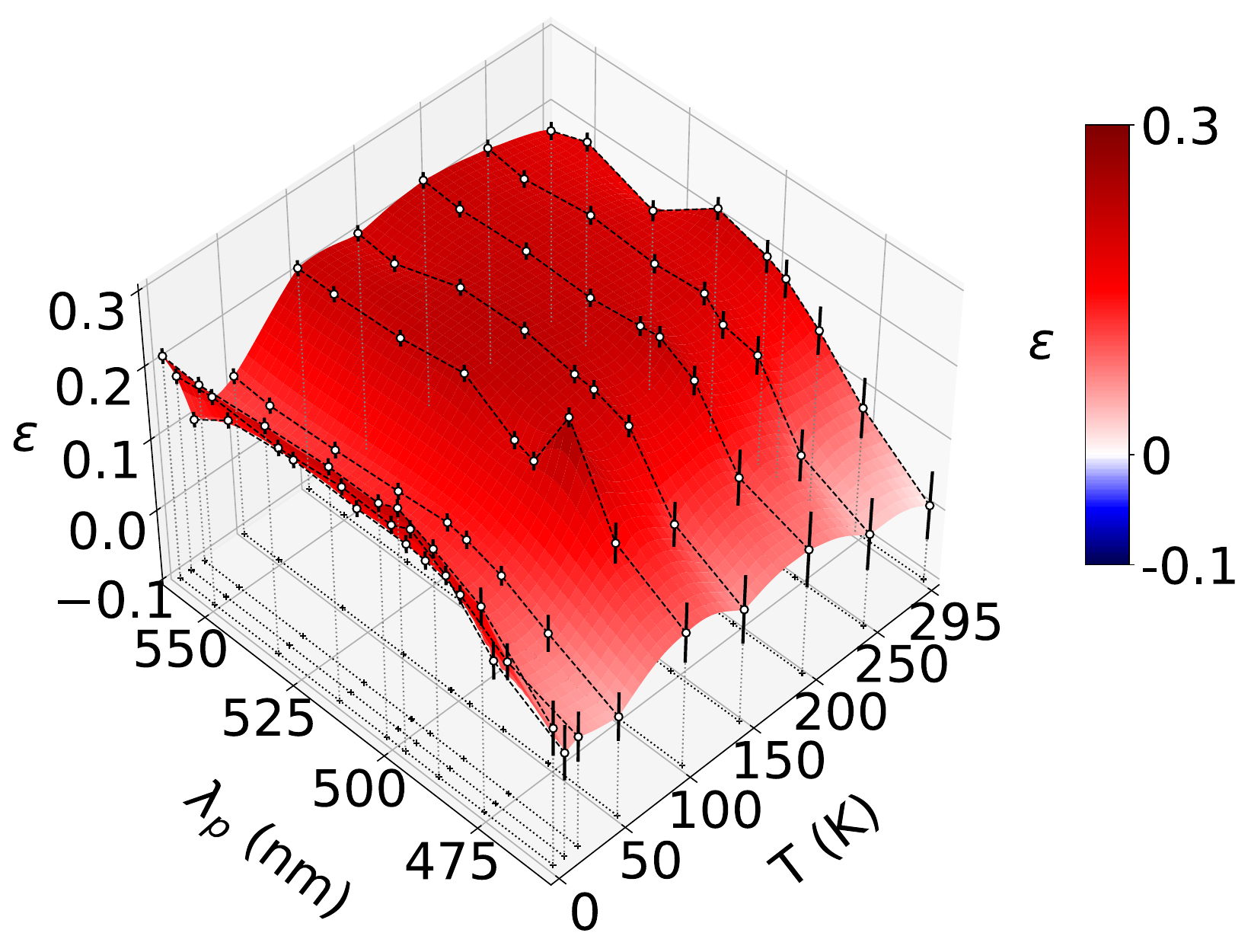}
    \caption{Field-induced contrast for temperatures between 1.6~K and 295~K and wavelengths between 455~nm and 560~nm. Positive values of $\epsilon$\ correspond to decreased PL amplitude.}
    \label{fig:eps_trench}
\end{figure}

Fig.\ \ref{fig:eps_trench} shows the contrast parameter $\epsilon$ which describes a reduction in \nvm\ PL with applied field as defined in (Eqns.\ 1, 2). The contrast $\epsilon$ shows a broad plateau between 100~K and 300~K bounded at shorter wavelengths by the ground state ionization threshold near 480~nm, and by a valley of suppressed contrast between 100~K and 1.6~K. Focusing first on the plateau, $\epsilon$ is expected to be independent of excitation wavelength in the low-power, linear-response regime as long as concentrations are not wavelength-dependent. The decrease below 480 nm coincides with depressed \nvm\ PL due to photoionization of the $^3A_2$ triplet state (shown in Fig.\ \ref{fig:minusfrac}).

\begin{figure*}[hbt!]
    \centering
    \includegraphics{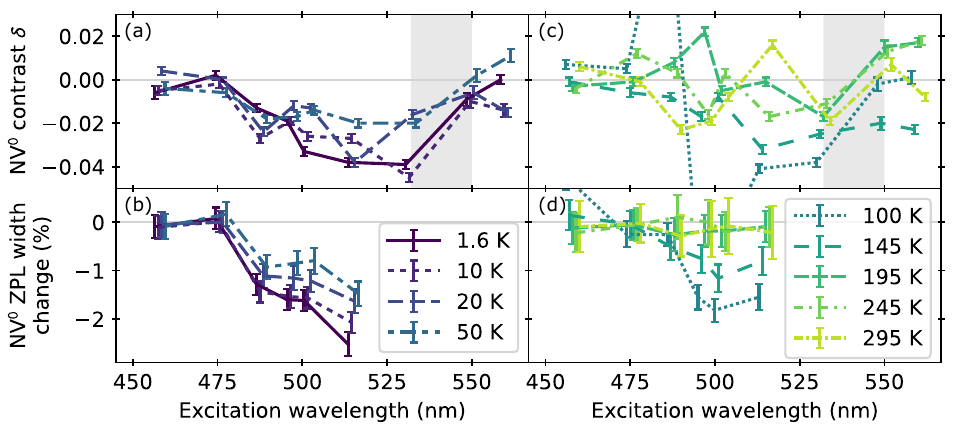}
    \caption{Field-induced contrast parameter ($\delta$). Field on/off contrast measurements performed below 100 K (a-b) and above 100 K (c-d) for (a), (c) error weighted power average of \nvz\ PL amplitude ($\delta$), and (b), (d) percent change in \nvz\ ZPL linewidth ($\alpha$) at 40 $\mathrm{\mu} W$ laser power for excitation wavelengths between 455 nm and 516 nm. Positive (negative) values of $\delta$\ correspond to decreased (increased) PL amplitude. Positive (negative) values of $\alpha$\ correspond to increased (decreased) ZPL \nvz\ linewidth.  Shaded regions in (a), (c) denotes approximate ionization threshold.}
    \label{fig:epsdelta}
\end{figure*}

The suppressed $\epsilon$ contrast shown in Fig.\ \ref{fig:epsdelta} around 50~K and across all wavelengths also appears in earlier results by Rogers et al.\cite{rogers_time-averaging_2009} and others by Ernst et al. and Happacher et al. contemporaneous with this work.\cite{ernst_temperature_2023,ernst2023modeling,happacher_low-temperature_2022} We propose an explanation for this effect related to thermal averaging of the excited state orbitals. At room temperature, the excited state behaves like a single spin triplet. However, at low temperature the $^3E$ excited state of \nvm\ is an orbital doublet which is split by transverse strain into $^3E_x$ and $^3E_y$ electronic states separated by tens of GHz.\cite{tamarat_spin-flip_2008, batalov_low_2009} At temperatures above 100~K, these spin triplets are thermally averaged, and the excited state appears as one triplet state.\cite{rogers_time-averaging_2009, fu_observation_2009, happacher_low-temperature_2022}. To explain the suppressed contrast, we propose that the thermal averaging process produces an effective field noise that causes fast relaxation of the spin states at intermediate temperatures. The spin relaxation rate will have a maximum when the characteristic switching time is at the GHz frequencies corresponding to transitions between the spin triplet sublevels. 
This spin-disordering mechanism will reduce polarization into the $m_z = 0$ state in zero field, so further mixing due to an applied field will produce a suppressed PL quenching effect.  Further mixing and PL quenching due to an applied field will have a suppressed effect. In support of this mechanism, we note that shallow minima in the zero-field \nvm\ PL fraction are visible in the same 50~K region as the suppressed contrast $\epsilon$.  See Fig.\ \ref{fig:minusfrac}(a). A thorough study of the contrast temperature dependence has recently been made public.\cite{ernst_temperature_2023}.

%
%

The \nvz\ has no known spin-dependent relaxation mechanisms, so its optical properties are considered immune to modest magnetic field. The contrast parameter $\delta$ is therefore an indicator of changes in \nvz\ concentration.  The field-immunity of \nvz\ also implies field-independent recombination mechanisms, so we attribute $\delta$ to changes in \nvz\ concentration induced by a field-dependent change in ionization rate of the \nvm. The plots of $\delta$ in Figs. \ref{fig:epsdelta}(a) and (c) show a region of negative contrast between approximately 480~nm and 540~nm, strongest below 100~K. Negative contrast here corresponds to an increase in \nvz\ PL induced by the applied magnetic field. 

As shown in Fig.\ \ref{fig:levels_and_fieldplot}(b), the population of the excited state decreases with applied field and the population of the singlet state sees a corresponding increase, strongly implying that the negative values of $\delta$ are due to ionization from the singlet state. This result adds to a growing body of experimental evidence \cite{hopper_near-infrared-assisted_2016, hacquebard_charge-state_2018, aude_craik_microwave-assisted_2020} indicating a wavelength dependent increase in \nvz\ population is a direct result of resonant CW and pulsed microwave excitation\cite{aude_craik_microwave-assisted_2020,hopper_near-infrared-assisted_2016, hacquebard_charge-state_2018} or DC field quenching (this work) inducing an increase in $\ket{\pm 1}$ spin state population.

Regions of negative $\delta$ in Figs.\ \ref{fig:epsdelta}(a) and (c) constitute the main results of this paper. The low wavelength boundary of the negative=$\delta$ region near 480~nm coincides with the abrupt dropoff in \nvm\ PL, which we interpret as the threshold for ionization from the ground state. We interpret the high wavelength boundary of the negative-$\delta$ region near 540~nm to be a threshold for ionization from the $^1E$ long-lived singlet state. The threshold appears between excitation wavelengths of 532~nm and 550~nm, bracketing the singlet ionization threshold between 2.25~eV and 2.33~eV.
At 200~K and above, the negative contrast weakens significantly or changes sign, indicating a change in ionization rate as a function of temperature that is likely a result of multiple competing processes, including decreased lifetime of the singlet state,\cite{acosta_optical_2010,robledo_spin_2011,thiering_theory_2018}, thermal occupation of vibronic states of the $^1A_1$ and $^1E$,\cite{dhungel_zero-field_2023} temperature dependent shifts in valence and conduction band energies,\cite{gali_recent_2023} and small shifts in the singlet state energies within the band gap.\cite{gali_recent_2023}

%
%

The increase in \nvz\ concentration indicated by negative values of $\delta$ suggests reactions that reduce the number of charged centers in the diamond. Ionization, ${\rm N}V^-\rightarrow{\rm N}V^0 + e^-$, neutralizes an \nv\ center, and if the freed electron then goes on to neutralize a \nsp\ ion, the total number of charged centers in the sample is reduced by 2. The accompanying reduction in random electric fields would be observable as a narrowing of ZPL peaks as Stark shift broadening is reduced.\cite{Orphal-Koban_spectraldiffusion2023}

Figs.\ \ref{fig:epsdelta}(b) and (d) show the fractional change in the width of the \nvz\ ZPL with applied field; negative values correspond to narrowing. The results shown in Figs.\ \ref{fig:epsdelta}(b) and (d)) include a wavelength region above 488~nm for temperatures below 150~K where the width of the \nvz\ ZPL line decreases by $\approx$2~\% with applied field, in agreement with the $\delta$ contrast for that region. This line width change and the $\delta$ contrast both point to significant ionization from the singlet state.

We do not attempt to explain the anomalous behavior at 100~K in Figs.\ \ref{fig:epsdelta}(c) and (d), which includes one sharp peak at certain wavelengths below 490~nm and adjacent pit above 490~nm. These anomalies appear consistently over different laser powers and the different wavelengths were measured on different days.

Surface plots of the results presented in Fig.\ \ref{fig:epsdelta} are included in the supplemental material for a synoptic overview of these results.

%% file: conclusion.tex
The \nv\ center in diamond has become a canonical quantum system, featured in work on fundamental quantum effects, quantum sensing and physics education. An energy level diagram similar to Fig.\ \ref{fig:levels_and_fieldplot}(a) appears in countless publications. So, it may be surprising that there are parts of the \nvm\ energy level diagrams that are not well-known. In particular, the energies of the singlet states relative to the triplet states have proven difficult to measure. In this work, we probed the ionization dynamics of \nvm\ centers using applied magnetic field modulation. We have identified ionization thresholds for both the $^3A_2$ ground state and the long-lived $^1E$ singlet state, and discovered a temperature and wavelength dependence of the ionization rate from the $^1E$ singlet, which has not been measured previously. Knowledge of ionization pathways will be important as new research in electrical readout and spin-charge conversion strives to circumvent the poor optical readout characteristics of \nv\ centers.

For temperatures between 2 K and 200 K, the ionization threshold of $^1E$ was found between 2.33~eV and 2.25~eV (excitation wavelengths 532~nm and 550~nm resp.), which is in reasonable agreement with 2.22~eV predicted by {\em ab\ initio} calculations.\cite{bockstedte_supplementary_2018} For temperatures above 200 K the ionization rate appears to change as a function of temperature. Ionization from the singlet state is identified by increases in \nvz\ PL with applied field, supported by narrowing of the ZPL linewidth due to the Stark effect. Combined with our observation of a ground state ionization threshold between 2.54~eV and 2.60~eV, we bracket the energy difference 0.21 eV $< \Sigma <$ 0.35 eV, which is in reasonable agreement with prior results inferring this value from values of $\Delta$. \cite{goldman_phonon-induced_2015,goldman_state-selective_2015,goldman_erratum_2017,thiering_theory_2018}.

%% file: analysis.tex
The photoluminiscence (PL) spectra are analyzed using Bayesian inference with a the goal of estimating parameters $\epsilon$ and $\delta$ from arrays of measurement data 
${\bm I_0} \equiv \{I_{0, 1}, I_{0, 2}, \ldots\}$, 
${\bm I}_B \equiv \{I_{B, 1}, I_{B, 2}, \ldots\}$, and wavelength 
${\bm \lambda} \equiv \{ \lambda_1, \lambda_2$, \ldots\}. $I_0$ and $I_B$ are spectra without and with magnetic field, respectively. Using the notation $P(A|B)$ to indicate the probability of $A$ given (or conditional on) $B$, Bayes' rule provides
\begin{equation}
    P(\epsilon, \delta | {\bm I}_{0}, {\bm I}_{B}, {\bm \lambda}) \propto
    P({\bm I_{0}}, {\bm I_{B}} | \epsilon, \delta, {\bm \lambda})\; P_0(\epsilon, \delta).
    \label{eq:bayesrule}
\end{equation}
The left side is the desired posterior probability distribution of parameters $\epsilon$ and $\delta$ given measurement data ${\bm I}_{0}, {\bm I}_{B}$ and ${\bm \lambda}$. The trailing term on the right side is the prior, which expresses any  knowledge of $\epsilon$ and $\delta$ values before considering the measurement data. We use a prior with $\epsilon$ uniformly distributed between -0.5 and 0.5 and $\delta$ uniformly distributed between -0.5 and 0.2, reflecting a typical reported PL contrast of 10 \% to 20 \% and a cursory examination of the data.

The first term on the right side of (\ref{eq:bayesrule}) is the likelihood, the probability of obtaining data ${\bm I}_0$ and ${\bm I}_B$ at wavelengths ${\bm \lambda}$ given parameters $\epsilon$ and $\delta$. In the current context, the likelihood is essentially a model of spectral data as a function of the parameters, expressed as a probability distribution.  To develop likelihood expressions, we construct a data model where the PL spectra are the a sum of contributions from centers in the \nvz and \nvm charge states plus a background and noise.
\begin{subequations} \label{eq:model}
\begin{eqnarray}
I_{0,n} & = & I^-_n + I^0_n + C + \eta_0
\label{eq:modela} \\
I_{B,n} & = & (1-\epsilon)I^-_n + (1-\delta)I^0_n + C + \eta_B.
\label{eq:modelb}
\end{eqnarray}
\end{subequations}
Measurement data $I_{0, n}$ and $I_{B,n}$ are spectrum values for wavelength $\lambda_n$ recorded with the magnetic field off and on, respectively. Contributions to $I_{0, n}$ from \nvm and \nvz centers are $I^-_n$ and $I^0_n$. Parameters $\epsilon$ and $\delta$ respectively quantify quenching of the $I^-_n$ and $I^0_n$ photoluminescence when the magnetic field $B$ is applied. Wavelength-independent $C$ is a background level, and random variables $\eta_0$ and $\eta_B$ represent zero-mean, Gaussian measurement noise with standard deviations $\sigma_0$ and $\sigma_B$, respectively.
    
For arbitrary index $n$, the two equations of (\ref{eq:model}) include two unknown values $I^-_n$ and $I^0_n$ in addition to unknown parameters $\epsilon$, $\delta$, and $C$. Progress can be made if inference is limited to a few zones of the spectrum where extra information is available. Figure \ref{fig:fitexample2} shows the zones in relation to a sample spectrum, and the extra information applied to each zone is described below.


\subsubsection*{Inference procedure}

\paragraph*{Zone BG}
The inference procedure starts by establishing the background in zone BG, which
is assumed to contain measurements of $C$ with no contribution from photoluminescence. The mean and variance of combined $I_{0, n}$ and $I_{B,n}$ data ($n \in$ BG) establish a normal (Gaussian) probability distribution $P_{BG}(C)$ which is used as a prior in the next stage. To simplify notation, we define
\begin{equation}
    P_{Z}(\cdot) = P(\cdot | I_{0, n}, I_{B, n}, n \in Z)
\end{equation}
as a parameter distribution conditional on (i.e. after inference using) all data from zone $Z$.

\paragraph*{Zone ZPL0}
Zone ZPL0 encompasses the narrow zero phonon line (ZPL) of the \nvz, which occurs at wavelengths well below the \nvm photoluminescence spectrum. In this zone, the analysis assumes that PL data has contributions from \nvz and background only, so $I^-(\lambda) = 0$. Information from this zone largely determines $\delta$.

Inference in zone ZPL0 requires a prior distribution for parameters $C$ and $\delta$ and an expression for likelihood. For the prior, we use the product $P_0(C, \delta) = P_{BG}(C)P_0(\delta)$, which includes the background $P_BG(C)$ and a uniform distribution $\delta \sim {\cal U}[-0.5, 0.2]$. 

Maniuplation of (\ref{eq:modela}) and (\ref{eq:modelb}) to eliminate the unknown $I^0_n$ yields
\begin{equation}
    (1-\delta)I_{0, n}(\lambda) - I_{B,n}(\lambda) + \delta C = (1-\delta)\eta_0 + \eta_B.
\end{equation}
The terms here are arranged so that the right side has known statistical properties. (In the larger picture, the $\delta$ parameter has an unknown distribution, but for the likelihood $\delta$ is one of the given values, so it is treated as a number, not a random variable here.) The expectation (mean) ${\cal E}[\cdot]$ and variance ${\cal V}[\cdot]$ of this expression are
\begin{widetext}
\begin{eqnarray}
    {\cal E}[(1-\delta)I_{0, n}(\lambda) - I_{B,n}(\lambda) + \delta C] & = & 0 \\ 
    {\cal V}[(1-\delta)I_{0, n}(\lambda) - I_{B,n}(\lambda) + \delta C] & = &(1-\delta)^2\sigma_0^2 + \sigma_B^2
\end{eqnarray}
where  $\sigma_0$ and $\sigma_B$ are standard deviations expressing uncertainty in measurement data $I_{B,n}$ and $I_{B,n}$ respectively.  The likelihood of measurement values $I_{B,n}$ and $I_{B,n}$ is a normal distribution

\begin{equation}
    P(I_{0,n}, I_{B,n} | \delta, C) \propto \exp\left[
    \frac{-[(1-\delta)I_{0, n} - I_{B,n} + \delta C]^2}
    {2[(1-\delta)^2\sigma_0^2 + \sigma_B^2]}\right]
\end{equation}
Application of Bayes' rule for all indices in the zone yields the posterior joint distribution $P_{\rm ZPL0}(\delta, C)$
\begin{equation}
    P_{\rm ZPL0}(\delta, C) \propto \prod_{n \in {\rm ZPL0}} 
    P(I_{0,n}, I_{B,n} | \delta, C)   \;\;P_0(C,\delta)
\end{equation}
\end{widetext}

\paragraph*{Zone ZPL-}
Zone ZPL- encompasses the narrow ZPL peak of the NV$^-$ spectrum, providing information to determine $\epsilon$. The signal in this zone includes contributions from the ZPL peak of \nvm and also from the NV$^0$ phonon sideband. At wavelengths within zone ZPL-, $I^0$ is expected to be relatively featureless. We effectively assign the peak to the $I^-(\lambda)$ contribution by modeling the \nvz contribution with as a linear function of $\lambda$, $I^0 \approx m (\lambda - \bar{\lambda}) + b$.\cite{aude_craik_microwave-assisted_2020,chakraborty_magnetic-field-assisted_2022}  Here, $\bar{\lambda}$ is chosen to be the wavelength at the center of the zone.  Manipulation of (\ref{eq:modela}) and (\ref{eq:modelb}) yields
\begin{widetext}
\begin{equation}
(1-\epsilon)I_{0, n} - I_{B,n} + (\epsilon - \delta) m (\lambda_n - \bar{\lambda})
+ (\epsilon - \delta)b + \epsilon C = (1-\epsilon)\eta_0 - \eta_B
\label{eq:zplm_likelihood}
\end{equation}
the effective slope, $(\epsilon - \delta)m$ and effective $C^* = (\epsilon - \delta)b + \epsilon C$. 
The expectation value and variance of this expression lead to a likelihood
\begin{equation}
    P(I_{0,n}, I_{B,n} | \epsilon, \delta, m^*, c^*)  \propto 
     \exp\left[
    \frac{-[(1-\epsilon)I_{0, n} - I_{B,n} + (\epsilon - \delta)m(\lambda_n - \bar{\lambda}) + (\epsilon - \delta)b + \epsilon C]^2}
            {2[(1-\epsilon)^2\sigma_0^2 + \sigma_B^2)]}
    \right]  
\end{equation}
\end{widetext}
\subsubsection*{Implementation}
We implement the probability distributions as a particle filter / sequential Monte Carlo method. The distribution is represented by a collection of points in parameter space, $\theta_n \equiv (\epsilon_n, \delta_n, m_n, c_n)$ with weights $w_n$.
\begin{equation}
    P(\theta) \approx \sum_n \delta(\theta - \theta_n) w_n.
\end{equation}


%% file: photodynamics.tex
The five-state model for the photodynamics of \nvm\ has been described by several authors.\cite{robledo_spin_2011, tetienne_magnetic-field-dependent_2012, liaugaudas_luminescence_2012, beha_optimum_2012, hacquebard_charge-state_2018} In this model the ground state triplet is represented by two states: a $S_z$ = 0 state and a combined $S_z = \pm 1$ state.  Similarly, the excited triplet is represented by two states and the fifth state is the singlet "shelving" state. Transitions include photon absorption from ground to excited states, photon emission from excited to ground states, strongly spin-dependent relaxation from excited states to the singlet state and slow, mildly spin-dependent relaxation from the singlet state to the ground state.

For computation, values for intrinsic relaxation rates were adopted from Tetienne et al.\cite{tetienne_magnetic-field-dependent_2012} The excitation cross sections are modeled using the form
\begin{equation}
    \sigma_g(E_{\rm phot}) = \sigma_0 e^{-(E_{\rm phot} - E_{\rm max})^2/(2w^2)},
\end{equation}
where $\sigma_0$ is a cross section, $E_{\rm max}$ is the photon energy at the maximum of the phonon band absorption and $w$ is a width parameter. The excitation cross section of \nvm\ is modeled using $\sigma_0 = 0.0045$~nm$^2$, $E_{\rm max}=2.17$~eV and $w = 0.21$~eV. These values are chosen to approximate the theoretical results of Razinkovas et al.\cite{razinkovas_photoionization_2021}. For \nvz\ the values are $\sigma_0 = 0.0045$~nm$^2$, $E_{\rm max}=2.28$~eV and $w = 0.21$~eV. These values shift the phonon band by 0.21~eV (the difference in zpl energies). The unchanged $\sigma_0$ value ensures that the \nvz\ cross section is 130~\% of the \nvm\ cross section under 532~nm illumination in agreement with measurements.\cite{hacquebard_charge-state_2018} The recombination cross section into the \nvz\ excited state was arbitrarily assumed to be the same as the ionization cross section from the \nvm\ excited state.